\documentclass[aps,prb,twocolumn,showpacs,superscriptaddress]{revtex4}
\usepackage{graphicx}
\usepackage{amsfonts,amssymb,amsmath}
\usepackage{bm}
 
\begin{document}
 
\title{Anisotropy of the orbital moments and the magnetic
dipole term $T_z$ in ${\rm CrO_2}$: 
An {\it ab-initio} study}

\date{\today}
 
\author{Matej Komelj}
\affiliation{Jo\v zef Stefan Institute, Jamova 39, SI-1000 Ljubljana,
  Slovenia}
\author{Claude Ederer}
\affiliation{Materials Research Laboratory, University of California,
  Santa Barbara, CA 93106-5130, U.S.A.}
\author{Manfred F\"ahnle}
\email{faehnle@mf.mpg.de}
\affiliation{Max-Planck Institut f\"ur Metallforschung, Heisenbergstr.~3,
  D-70569 Stuttgart, Germany}

\begin{abstract}
A systematic study is performed by the {\it ab-initio} density
functional theory of the anisotropy of the orbital moments 
$\langle l_z\rangle$ and the
magnetic dipole term $\langle T_z\rangle$ in bulk ${\rm CrO_2}$. Two
different band-structure techniques are used (FLAPW and LMTO-ASA), and
the electronic correlations are treated by the local-spin-density
approximation (LSDA), the LSDA+ orbital polarization method, and the
LSDA+$U$ method.  The calculated anisotropies of $\langle l_z\rangle$
and $\langle T_z\rangle$ are very large compared to Fe, Ni and Co but
still a factor of 5 and 2 smaller than the anisotropies obtained from
a recently suggested analysis of the X-ray magnetic circular dichroism
spectra for a thick layer of ${\rm CrO_2}$.
\end{abstract}

 
\maketitle

Recent research within the fields of magnetic tunneling and spin
injection involves ${\rm CrO_2}$ as a promising material for
electrodes \cite{Julliere:75}. Besides its potential importance for
applications, this material also exhibits interesting physics
originating from its half metallic nature and consequently the
ferromagnetism due to the double-exchange coupling \cite{Korotin:98},
as well as from its orbital magnetism\cite{Goering:02,Huang:02} which is
related to the spin-orbit coupling and the electronic orbital
correlation effects.  Consequently, there are already several
experimental \cite{Goering:02,Huang:02} and theoretical \cite{Kunes:02,
Uspenskii:96,Oppeneer:01,Huang:02} investigations of the orbital
moments $\langle l_z\rangle$ in ${\rm CrO_2}$. \par However, much less
information is available on the anisotropy of the orbital moments,
i.e., on its dependence of the orientation of the sample magnetization
(in the following the $z$ axis of the external coordinate system is
always chosen to be parallel to the magnetization direction). It has
been pointed out first by Bruno\cite{Bruno:89} and later worked out in
more detail by van der Laan \cite{vanderLaan:98} that the anisotropy
of the orbital moment is closely related to the magnetocrystalline
anisotropy energy. Furthermore, van der Laan has shown that there is
an additional contribution to the anisotropy energy arising from the
anisotropy of the magnetic dipole term $T_z$, which is the expectation
value of the magnetic dipole operator:
\begin{equation}
\hat{T}_{z} = \frac{1}{2} [ \bm{\sigma} -
3\hat{\mathbf{r}}(\hat{\mathbf{r}}\cdot\bm{\sigma}) ]_{z}.
\end{equation}
In eq.(1) $\hat{\mathbf{r}}$ is the unit vector in the direction of
the position vector $\mathbf{r}$, and $\bm{\sigma}$ is the vector of
the Pauli matrices.  In fact, it turns out (see below) that for ${\rm
CrO_2}$ the two contributions are very large and of similar magnitudes
but opposite in sign.\par A suitable method to investigate the
anisotropy of $\langle l_z\rangle$ and $\langle T_z\rangle$ is the
angle-resolved variant \cite{Stoehr/Koenig:1995} of the X-ray magnetic
circular dichroism \cite{Schuetz:1987} (XMCD) which measures the XMCD
spectra for various orientations of the sample magnetization.  The
orbital moments $\langle l_z\rangle$ then may be determined directly
from the application of the XMCD orbital sum rule\cite{Thole:1992}.
In contrast, the application of the spin sum rule \cite{Carra:1993}
yields a combination of the spin moment $\langle\sigma_z\rangle$ and
of the $\langle T_z\rangle$ term. A separation of these two
contributions is possible by applying the spin sum rule to XMCD
spectra for various orientations of the sample magnetization. A
precondition of the use of this method is that $\sum_\alpha\langle
T_z\rangle_\alpha\approx 0$, where $\alpha$ represents one of the
perpendicular directions in which the $z$ axis is oriented. It has
been outlined in Ref. \onlinecite{Ederer:2003} that the validity of
this relation is not always guaranteed (for the case of ${\rm CrO_2}$
see below).  \par
Goering {\it et al.} \cite{Goering:02} applied the
technique of angle-resolved XMCD to investigate the anisotropy of
$\langle l_z\rangle$ and $\langle T_z\rangle$ for the Cr atom in a
thick layer of ${\rm CrO_2}$ on a ${\rm TiO_2}$ substrate (which is
required to stabilize the rutile structure).  For ${\rm CrO_2}$ the
analysis of the XMCD spectra of the $L_2$ and $L_3$ edges of the Cr
atom (corresponding to the $2p_{1/2}\rightarrow 3d$ and
$2p_{3/2}\rightarrow 3d$ transitions) is more difficult because the
spin-orbit coupling of the Cr $2p$ electrons is relatively weak so that
the $2p_{1/2}$ and $2p_{3/2}$ levels are not well
separated\cite{Scherz:02}. The subdivision of the XMCD spectra into the
$L_2$ and $L_3$ contributions (which is required for an application of
the spin sum rule) is therefore highly problematic due to a possible
quantum-mechanical mixture of the $2p_{1/2}$ and $2p_{3/2}$ levels
and/or due to a strong overlap of the two contributions on the energy
scale.  In their original analysis Goering {\it et
al.}\cite{Goering:02} neglected the quantum-mechanical
mixture. Furthermore, they separated the $L_2$ and $L_3$
contributions by an empirical (and theoretical not yet justified)
extension of the van der Laan's method of the moment
analysis\cite{vanderLaan:97} (which was proposed for systems with very
small crystal-field splitting of the valence states) to situations with
large band splitting. Based on these two assumptions, they found
extremely large anisotropies of $\langle l_z\rangle$ and $\langle
T_z\rangle$. For instance, the difference in the orbital moment for
the magnetization along the $c$ axis and along the $a$ axis of the
rutile structure was $0.083\>\mu_\text{B}$, which is extremely large
compared to bulk materials with cubic symmetry
($10^{-4}\>\mu_\text{B}$, see Ref. \onlinecite{Wilhelm:00}) and
comparable to the anisotropy found for extremely thin Co
layers\cite{Weller95}.  The anisotropy of $\langle T_z\rangle$
appeared to be about $0.2\>\mu_\text{B}$, two orders of
magnitude larger than the absolute value of $\langle T_z\rangle$ in Fe
($0.004\>\mu_\text{B}$), Ni ($-0.004\>\mu_\text{B}$) and Co
($0.002\>\mu_\text{B}$)\cite{Wu:1994-1}. For the true spin moment of the
Cr atom a value of $1.2\>\mu_\text{B}$ was obtained. To explain the
magnetic moment of about $2\>\mu_\text{B}$ per unit cell which was obtained 
by a SQUID measurement Goering {\it et al.}\cite{Goering:02} assumed a very 
large
spin moment of about $0.4\>\mu_\text{B}$ per O atom which they tried to
explain in terms of a hybridization between chromium and oxygen. \par
The question is whether the extreme anisotropies of $\langle
l_z\rangle$ and $\langle T_z\rangle$ found by Goering {\it et
al.}\cite{Goering:02} are indeed intrinsic properties of bulk ${\rm
CrO_2}$ with rutile structure. Remember that their measurements were
performed on thick layers of ${\rm CrO_2}$ on ${\rm TiO_2}$. So far
there is only very little information on the structure of this layer,
i.e., on possible oxygen deficiencies. Furthermore, it is not known
how the two basic assumptions for the application of the spin sum rule
(see above) affect the results for $\langle T_z\rangle$ and for the
true spin moment $\ \langle\sigma_z\rangle$. Indeed, most recently
\cite{Goering:private} an empirical method has been developed to take
into account approximately the effect of mixing of the $p_{1/2}$ and
$p_{3/2}$ core states for the analysis of the XMCD spectra by the spin
sum rule. The data of Goering {\it et al.} were
reanalyzed\cite{Goering:private} by this method, yielding a Cr moment
that agreed reasonably well with the Cr moment suggested by the SQUID
measurements after neglecting the O moment, but the anisotropy of the
$\langle T_z\rangle$ remained to be very large when obtained by the
help of the above discussed moment analysis. \par In the present paper
we investigate theoretically the anisotropy of $\langle l_z\rangle$
and $\langle T_z\rangle$ in bulk ${\rm CrO_2}$ with rutile structure
by calculations based on the {\it ab-initio} density functional
electron theory. In the literature there are already some single
results in this direction which -- however -- are in part contradictory
\cite{Kunes:02,Uspenskii:96,Oppeneer:01}. We therefore have performed
a systematic study, which is based on two different band-structure
techniques and which takes a special care of the electronic orbital
correlation effects (again by two different methods) which are not
included in the commonly used local-spin-density approximation (LSDA)
of the density functional theory.

We performed comparative calculations by two different band structure
methods, the Wien97 code \cite{Blaha:1990}, which adopts the
full-potential linearized-augmented-plane-wave
method\cite{Wimmer:1981} (FLAPW) and the linear-muffin-tin-orbital
method in the atomic-sphere approximation \cite{Andersen:75}
(LMTO-ASA). Whereas in the first method the effective crystal
potential is treated exactly, it is spherically averaged in each
atomic sphere after each iteration step of the self-consistency cycle
in the LMTO-ASA method. The exchange-correlation potential was
calculated in LSDA \cite{Perdew:1992}. The electronic orbital
correlation effects, which are not included in LSDA, were taken into
account by the OP term\cite{Eriksson:1990} or by means of the LDA+$U$
scheme\cite{Anisimov:93,Liechtenstein:95}.  For $U=0$, the LDA+$U$
calculation is equivalent to a LSDA calculation. In the case of
$U\ne0$ the exchange-interaction parameter $J$, which appears in the
LDA+$U$ scheme in addition to the parameter $U$, was fixed\cite{Korotin:98} to
$0.87\>{\rm eV}$, whereas $U$ was an open parameter
(the calculated\cite{Korotin:98} screened value of $U$ for Cr in ${\rm
CrO_2}$ is $U=3\>{\rm eV}$). In the literature arguments for and
against the use of the LDA+$U$ method for the case of ${\rm CrO_2}$
are given (see, e.g., Ref.  \onlinecite{Kunes:02} and
Refs. therein). Therefore we performed a comparative study based on
various calculational schemes for the electronic correlations.  The
spin-orbit coupling, OP term and LDA+$U$ scheme were implemented in
the LMTO code in Ref. \onlinecite{EdererPhd}. The magnetization was
set along the $c$ or the $a$ axis of the rutile structure. The
calculations were performed for the experimental lattice 
parameters\cite{Porta:72}
$a=0.4419\>{\rm nm}$, $c=0.2912\>{\rm nm}$ and
$u=0.303$.

\begin{table*}
\begin{ruledtabular}
\begin{tabular}{llrrrrrr}
&$U[{\rm eV}]$	  &0      &2       &3       &4       &5          & LSDA+OP\\
\hline
$c$ axis&$\langle l_z\rangle$      &-0.037 (-0.055)&-0.053& -0.064 (-0.090)& -0.083& -0.117 &   -0.054\\
Cr&$\langle\sigma_z\rangle$	  &1.89 (2.04)& 1.97&  2.04 (2.03)&  2.09&  2.12 &    1.89\\
&$\langle T_z\rangle$       &-0.059 (-0.048)& -0.062& -0.066 (-0.044)& -0.072& -0.084 &  -0.061\\
\hline
O&$\langle l_z\rangle$      &-0.0012 (-0.0014)&-0.0019& -0.0025 (-0.0018)& -0.0034 & -0.0051 &    -0.0018\\
&$\langle\sigma_z\rangle$      &-0.046 (-0.038)&-0.080& -0.114 (-0.036)& -0.131 & -0.145 &    -0.046\\
\hline\hline
$a$ axis&$\langle l_z\rangle$       &-0.035 (-0.045)&-0.044& -0.051 (-0.076)& -0.053& -0.061 & -0.049\\
Cr&$\langle\sigma_z\rangle$        &1.89 (2.04)&  1.96&  2.00 (2.03)&  2.10&  2.14 &  1.89\\
&$\langle T_z\rangle$       &0.031 (0.024)&0.040& 0.046 (0.022)& 0.043& 0.043 &0.031\\
\hline
O&$\langle l_z\rangle$	   &-0.0001 (0.0000)&-0.0001& -0.0008 (-.0001)&  -0.0000&  -0.0002 &  -0.0003\\
&$\langle\sigma_z\rangle$        &-0.046 (-0.038)&-0.077& -0.107 (-0.035)&  -0.139& -0.155 &  -0.046\\
\hline\hline
Cr&$\left|\Delta\langle l_z\rangle\right|$&0.002 (0.010)&0.009&0.013 (0.014)&0.03&0.056 &0.005\\
&$\left|\Delta\langle T_z\rangle\right|$&0.09  (0.072)&0.102&0.112 (0.066)&0.115&0.127 &0.092\\
\end{tabular}
\end{ruledtabular}
\caption{The calculated  values of $\langle l_z\rangle$, 
$\langle\sigma_z\rangle$, $\langle T_z\rangle$ and the anisotropies
$\left|\Delta\langle l_z\rangle\right|$ and $\left|\Delta\langle T_z\rangle\right|$,
all in $\mu_\text{B}$, from the LSDA calculations ($U=0\>{\rm eV}$), the LDA+$U$ 
calculations for various $U$ and for the LSDA+OP calculations. The values in brackets
are from the LMTO-ASA calculations, all other values are from the FLAPW
calculations.}
\end{table*}
The results are presented in Table 1. Hund's third rule is obeyed for both
magnetization directions at any value of $U$ because the orbital moments are
always antiparallel (parallel) to the spin moments at the Cr(O) sites, in 
agreement with the experimental observations \cite{Goering:02,Huang:02}. \par
We first discuss the orbital moments. From the LSDA calculation, the
LSDA+OP calculation and for the LDA+$U$ calculation with $U\le 3\>{\rm eV}$ we
find Cr orbital moments of about $-0.05\>\mu_\text{B}$ which magnitudes are 
smaller than
the experimental values\cite{Chen_et_al:1995} for Fe ($0.086\>\mu_\text{B}$) and
Co ($0.153\>\mu_\text{B}$). For the magnetization along the $c$ axis
our FLAPW calculation with $U=3\>{\rm eV}$  yields orbital moments of
$-0.064\>\mu_\text{B}$ and $-0.0025\>\mu_\text{B}$ at Cr and O sites, in a
very good agreement  with the XMCD results of Ref. \onlinecite{Huang:02} which
give $-(0.06\pm 0.02)\>\mu_\text{B}$ and $-(0.003\pm 0.001)\>\mu_\text{B}$, 
respectively. We take this as a hint that $U=3\>{\rm eV}$ (which is also the 
calculated\cite{Korotin:98} value for the screened $U$)  is a good choice. 
This is further underpinned by the fact that our LSDA+OP calculation yields
very similar results for the orbital moments of chromium and oxygen as
the LDA+$U$ calculation for $U$ between $2$ and $3\>{\rm eV}$. For
$U\le 3\>{\rm eV}$ our FLAPW results for the Cr (O) orbital moment agree
very well (perfectly) with those of a full-potential LMTO (FP-LMTO)
calculation\cite{Huang:02}. The probably less-accurate LMTO-ASA calculation 
yields larger Cr orbital moments than the FLAPW calculation. For $U=3\>{\rm eV}$
the LMTO calculation gives a value of $-0.090\>\mu_\text{B}$ for the $c$-axis 
orientation 
which agrees very well with the XMCD result of Ref. \onlinecite{Goering:02}. 
For the magnetization along the $a$ axis (for which there is no FP-LMTO
result in Ref. \onlinecite{Huang:02})  both our FLAPW and our LMTO-ASA
calculations yield smaller orbital moments than those for the 
$c$-axis orientation, in agreement with the XMCD results\cite{Goering:02}. 
The difference $\left|\Delta \langle l_z\rangle\right|=\left|\langle l_z
\rangle_\text{c axis}-\langle l_z\rangle_\text{a axis}\right|$ is 
about $0.015\>\mu_\text{B}$ at $U=3\>{\rm eV}$, which is indeed very large in
comparison to the orbital-moment anisotropy of cubic bulk materials ($10^{-4}\>
\mu_\text{B}$), but still a factor of about 5 smaller than the orbital-moment
anisotropy of ${\rm CrO_2}$ discussed by Goering {\it et al.}\cite{Goering:02}.
In the FLAPW calculation the value of $\left|\Delta\langle l_z\rangle\right|$
decreases drastically when reducing $U$ and it amounts to $0.002\>\mu_\text{B}$
for the LSDA calculation (the FLAPW calculation\cite{Kunes:02}
based on the generalized-gradient approximation (GGA) gives an orbital moment
anisotropy of $0.001\>\mu_\text{B}$). In the LMTO-ASA calculation the 
reduction with decreasing $U$ is less dramatic: for $U=0\>{\rm eV}$ we find
$\left|\Delta\langle l_z\rangle\right|=0.01\>\mu_\text{B}$ (which is a factor
of about 5 smaller than the value found by a LMTO-ASA calculation by
Uspenskii {\it et al.}\cite{Uspenskii:96}). Increasing $U$ beyond 
$3\>\mu_\text{B}$ enhances the orbital moment anisotropy to the values 
comparable to $0.083\>\mu_\text{B}$ discussed by Goering {\it et al.}
\cite{Goering:02}, but we doubt that such large values of $U$ are physically
reasonable for the Cr atom in ${\rm CrO_2}$. \par
The magnetic dipole term $\langle T_z\rangle$ at the Cr atom depends only
slightly on the way we take into account the correlation-effects. The LSDA,
LSDA+OP and LDA+$U$ ($U\le 3\>{\rm eV}$) calculations with the FLAPW
method yield values of about $-0.06\>\mu_\text{B}$ ($+0.04\>\mu_\text{B}$) for
the $c$-axis ($a$-axis) orientation. These values are large compared to the 
LSDA values\cite{Wu:1994-1} for
 Fe ($0.004\>\mu_\text{B}$), Ni ($-0.004\>\mu_\text{B}$) and
Co ($0.002\>\mu_\text{B}$), and they arise from a large crystal-field
anisotropy. Such large values usually appear in systems with strongly 
reduced dimensionality\cite{Ederer:2003}, and they show that the $T_z$-term
cannot be neglected in the analysis of the XMCD spectra via the spin sum rule
\cite{Carra:1993} when one wants to arrive at realistic values for the true spin
moment. The $\langle T_z\rangle$ values from the LMTO-ASA calculation are a bit
smaller which results from the fact \cite{Ederer:2002} that in this method 
the spin and
charge densities are calculated for an effective potential, which is spherically
averaged in each atomic sphere. The anisotropy $\left|\Delta\langle T_z\rangle
\right|$ is very large, about $0.1\>\mu_\text{B}$, in the FLAPW calculation but
still a factor of 3 smaller than the one discussed by Goering {\it et al.}
\cite{Goering:02}. \par
Finally, we have checked by the FLAPW method the applicability of the 
angle-resolved spin sum rule analysis for the case of ${\rm CrO_2}$. 
This analysis is appropriate if $7\sum_\alpha\langle T_z\rangle_\alpha/
\langle\sigma_z\rangle\ll 1$. For $U=3\>{\rm eV}$ ($0\>{\rm eV}$) we find 
a ratio of $0.09$ ($0.01)$). \par
Based on our values for $\left|\Delta l_z\right|$ and $\left|\Delta T_z\right|$
we calculated according to Ref. \onlinecite{vanderLaan:98} the respective
contributions to the magnetocrystalline anisotropy energy. They appear to be
very large, of the same order of magnitude but opposite in sign. Because
small uncertainties in either of the two contributions induce very large
errors for the magnetocrystalline anisotropy, it does not make sense 
to calculate the latter quantity from our data. \par
Finally, we consider the results for the spin moments. In all our different
types of calculations we find a Cr spin moment which is only very slightly 
anisotropic and which exhibits a magnitude of about $2\>\mu_\text{B}$, the 
value predicted by Hund's first rule for the case of a ${\rm Cr^{3+}}$ ion 
but considerably larger than the value of $1.2\>\mu_\text{B}$ found by 
the special type of analysis of the XMCD data\cite{Goering:02} discussed above.
For the O atom our FLAPW calculation yields for $U=3\>{\rm eV}$
a spin moment of about $0.1\>\mu_\text{B}$ which is a factor of 4 smaller
than the one obtained by Goering {\it et al.}\cite{Goering:02} with their
special analysis of the XMCD results. Our FLAPW spin moments for oxygen
are a bit larger than those from the FP-LMTO calculation of Ref. 
\onlinecite{Huang:02}. Our LMTO-ASA data are even smaller than the FP-LMTO
values. This probably results from the fact that sizes of the spheres over which
the spin density is integrated are different in various calculations. 
The atomic sphere in  the LMTO-ASA calculation is larger than the muffin-tin
spheres in the FP-LMTO and FLAPW calculations and probably contains 
already a part of the Cr spin-density, which has an opposite sign.

To conclude, we performed a systematic investigation by the FLAPW and the LMTO-ASA
methods for the anisotropies of the orbital moment $\langle
l_z\rangle$ and the magnetic dipole term $\langle T_z\rangle$ in ${\rm
CrO_2}$. The electronic correlation effects were taken into account by
three different methods, the LSDA, LSDA+OP and the LDA+$U$ method. The
calculated anisotropies of $\langle l_z\rangle$ ( $\langle
T_z\rangle$) are very large but still a factor of 5 (2) smaller than
those obtained by Goering {\it et al.}  \cite{Goering:02} from their
special type of analysis of XMCD data for thick layers of ${\rm
CrO_2}$ on ${\rm TiO_2}$. To arrive at such large anisotropies we had
to insert values for the parameter $U$ of the LDA+$U$ method which are
considerably larger than $3\>{\rm eV}$ and which seem to be physically
unreasonable. The only chance to arrive at larger anisotropies might
be to abandon the mean-field approximation  which is adopted in the
LDA+$U$ method for the electronic on-site correlations and to perform
a calculation within the dynamical mean field theory
(DMFT)\cite{Georges:96}. \par On the experimental side, more
investigations are required on the real structure of the considered
layers as well as on the way how to analyze the XMCD spectra which
arise from energetically very close core levels.\par Acknowledgement:
The authors are indebted to E. Goering, G.Y. Guo and M. Korotin for
helpful discussion. The work at Santa Barbara was supported by the MRL
Program of the National Science Foundation under Award
No. DMR00-80034.  
\bibliography{cro2.bib}

\end{document}